\documentclass[conference]{IEEEtran}
\IEEEoverridecommandlockouts
\usepackage{cite}
\usepackage{amsmath,amssymb,amsfonts}
\usepackage{algorithmic}
\usepackage{graphicx}
\usepackage{textcomp}
\usepackage{xcolor}
\def\BibTeX{{\rm B\kern-.05em{\sc i\kern-.025em b}\kern-.08em
    T\kern-.1667em\lower.7ex\hbox{E}\kern-.125emX}}
\begin{document}

\title{Interpretable simultaneous localization of MRI corpus callosum and classification of atypical Parkinsonian disorders using YOLOv5\\
}
\author{\IEEEauthorblockN{1\textsuperscript{st} Vamshi Krishna Kancharla}
\IEEEauthorblockA{
\textit{IIIT - Bangalore}\\
Karnataka, India \\
}
\and
\IEEEauthorblockN{2\textsuperscript{nd} Debanjali Bhattacharya}
\IEEEauthorblockA{
\textit{IIIT - Bangalore}\\
Karnataka, India \\
}
\and
\IEEEauthorblockN{3\textsuperscript{rd} Neelam Sinha}
\IEEEauthorblockA{
\textit{IIIT - Bangalore}\\
Karnataka, India \\
}
\\
\and

\IEEEauthorblockN{4\textsuperscript{th} Jitender Saini}
\IEEEauthorblockA{
\textit{Neuroimaging and Interventional Radiology} \\
\textit{NIMHANS - Bangalore}\\
Karnataka, India \\
}

\and
\IEEEauthorblockN{5\textsuperscript{th} Pramod Kumar Pal}
\IEEEauthorblockA{
\textit{Dept. of Neurology} \\
\textit{NIMHANS - Bangalore}\\
Karnataka, India \\
}
\and
\IEEEauthorblockN{6\textsuperscript{th} Sandhya M.}
\IEEEauthorblockA{
\textit{Neuroimaging and Interventional Radiology} \\
\textit{NIMHANS - Bangalore}\\
Karnataka, India \\
}
}

\maketitle

\begin{abstract}
Structural MRI(S-MRI) is one of the most versatile imaging modality that revolutionized the anatomical study of brain in past decades. The corpus callosum (CC) is the principal white matter fibre tract, enabling all kinds of inter-hemispheric communication. Thus, subtle changes in CC might be associated with various neurological disorders. The present work proposes the potential of YOLOv5-based CC detection framework to differentiate atypical Parkinsonian disorders (PD) from healthy controls (HC). With 3 rounds of hold-out validation, mean classification accuracy of 92\% is obtained using the proposed method on a proprietary dataset consisting of 20 healthy subjects and 20 cases of APDs, with an improvement of 5\% over SOTA methods (CC morphometry and visual texture analysis) that used the same dataset. Subsequently, in order to incorporate the explainability of YOLO predictions, Eigen CAM based heatmap is generated for identifying the most important sub-region in CC that leads to the classification. The result of Eigen CAM showed CC mid-body as the most distinguishable sub-region in classifying APDs and HC, which is in-line with SOTA methodologies and the current prevalent understanding in medicine.
\end{abstract}


\begin{figure*}[htbp]
\centerline{\includegraphics[width=15.5cm]{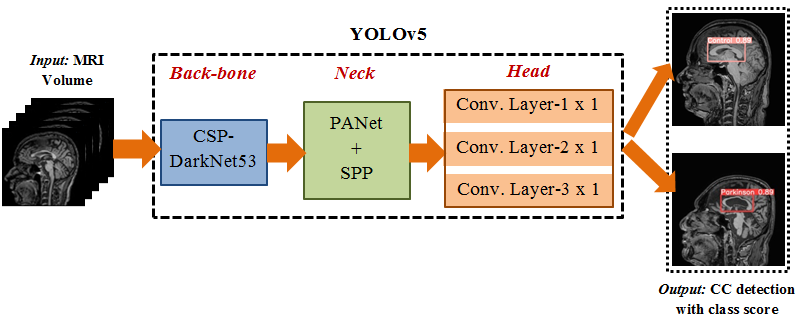}}
\caption{Block schematic for simultaneous detection of brain corpus callosum and classification of APDs vs. healthy control using YOLOv5 architechture}
\label{blockdiagram}
\end{figure*}

\section{Introduction}
Parkinson’s disease (PD) is a slowly progressing neurodegenerative movement disorder that mainly affects the motor system. The common motor and non-motor symptoms of PD include an ongoing loss of motor control (example, shaking, stiffness, slow movement, postural instability), loss of sense of smell, cognitive changes and many others. The category of Parkinsonism includes the classic form of PD and its other atypical variants, commonly referred “Parkinson’s Plus Syndromes” or “Atypical Parkinsonism”. These Atypical PDs (APD) are more severe and generally spread more rapidly; making it less treatable as compared to classical PD. Multiple System Atrophy (MSA) and Progressive Supranuclear Palsy (PSP) are two examples of such APDs, which are included in this study. Both PSP and MSA highlight the classical features of PD with additional rare symptoms that distinguish them from simple PD; although the clinical differentiation of PD, MSA and PSP at the early stage is extremely difficult \cite{b1,b2,b3}. \par
While there exists several literature that have studied the whole brain and regional brain atrophy to quantify neurodegeneration, little works have been performed on brain corpus callosum (CC) especially in case of APDs, despite its high relevance in neurodegeneration \cite{b4,b5,b6,b8,b9,b11}. The CC is the principal and the largest WM fibre track, situated at the center part of CNS. With more than 300 million fibres it connects two cerebral hemispheres and enabling all kinds of interhemispheric communication \cite{b14,b15}. Therefore, subtle structural changes in callosal architechture can be associated with cognitive and behavioral deficits seen in neurological diseases. The CC structure in brain is shown in Figure~\ref{cc}. Being easily distinguishable in mid-sagittal MRI plane, CC in this MRI plane lends itself well for extensive research \cite{b1,b2,b3,b4,b5,b6,b8,b9,b11}. Hence, in this work instead of analyzing changes in different sub-cortical brain regions, we focus in studying the degeneration of this particular CC structure in brain, seen in Parkinsonian disorders. \par
The present work is an extension of our previously published works that have quantified callosal degeneration in Parkinsonian disorders using CC morphometry and texture analysis \cite{b1,b2,b3,b4}. However, one major limitation of our previous studies was it relied on manual extraction of CC from MRI images. With advances in deep learning, some recent studies have used deep neural architecture for automatically segment CC structures from MRI image volume \cite{b17,b19,b20}. However, no studies have been reported on automatic detection of CC to classify Parkinsonian disorders from healthy controls (HC). This motivates us to examine the ability of deep learning based YOLO (You Only Look Once) framework to localize CC region in MRI volume while simultaneously classifying Parkinsonian disorders from controls. However, interpretibility of the results, which is crucial specially for medical images, needs to be illustrated. Since the differences in CC tissue alteration between HC and disease group may not be visually perceptible, in addition to CC detection and disease classification with respect to HC, class activation map (CAM) is built subsequently on MRI images in order to interpret the result of classification. This will in-turn aid to visually identify the most important callosal sub-region, leading towards classification. Thus, the key contribution of this work is to use YOLOv5 framework for simultaneous localization of brain CC and classification of APDs from HC, in combination with Eigen CAM in order to interpret the result of classification. 

\section{Dataset Description}

\begin{figure}[htbp]
\centerline{\includegraphics[width=6.5cm]{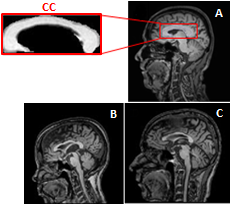}}
\caption{Representative MRI sagittal images of (A) healthy subject, (B) Multiple system atrophy and (C) Progressive supraneuclear palsy. The 'C-shaped' corpus callosum structure is shown in the left most column.}
\label{cc}
\end{figure}

The dataset for this study is collected from the general OPD and movement disorder services at Neurology department of National Institute of Mental Health and Neuro Science (NIMHANS), Bangalore. Dataset includes a total of 40 subjects: 20 patients with APDs: 10 patients with with clinically probable MSA (mean age: $53.90 \pm 5.53$) and 10 patients having with probable PSP an age: $63.50 \pm 7.36$). The MSA diagnosis was performed on the basis of Gilman criteria and PSP diagnosis was performed on the basis of criteria as illustrated in National Institute of Neurological Disorders and Stroke and Society. Twenty HC volunteers were also recruited ($55.3\pm 2$) and were age and gender matched, having no history of PD or other brain disorders. The study is approved by local ethics committee of NIMHANS. T1-W MRI was performed for all the subjects. Images were acquired in NIMHANS using a 3T MR system (Achieva; Philips Medical Systems, Netherlands) with a 32-channel head coil. Figure\ref{cc} shows representative MRI images of HC and patients suffering from MSA and PSP. Further details on subject demographics and MRI scan acquisition can be found in \cite{b1}. Each of the anatomical MRI scan was co-registered and normalized into a standard MNI space using SPM8 software.

\section{Proposed methodology}

This section describes YOLOv5 framework which is used in combination of Eigen CAM in order to simultaneously detect CC and classifying APDs from HC.

\subsection{YOLOv5 network architechture}
\label{yolov5}
In order to detect CC, YOLOv5 model is utilized \cite{b23,b24,b25}. The general architecture of YOLOv5 is shown in Figure~\ref{blockdiagram}. The YOLOv5 network architecture includes three main components: (i) Backbone, (ii) Neck and (iii) Head for prediction. The Backbone module consists of CSP-Darknet53 for feature extraction from images. This is the traditional CNN Darknet53 in which cross stage partial (CSP) network strategy is applied, that helps to truncate the gradient flow for reducing the large amount of redundant gradient information while preserving the advantage of feature reuse characteristics of DenseNet. Neck of YOLOv5 consists of modified path aggregation network (PANet) and spatial pyramid pooling (SPP). PANet is a feature pyramid network that helps to improve information flow and correct localization of pixels. In YOLOv5 model, CSPNet strategy is included in PANet. SPP performs information aggregation that it receives. Thus SPP has the advantage of significantly higher receptive field and separating majority of the relevant context features while maintaining the high speed of the network. The head of YOLOv5 consists of 3 convolution layers that predicts (i) the location of the bounding boxes, (ii) the object classes and (iii) the confidence-score of class prediction.

\subsection{Training methodology}
One important characteristics of PD is such that it distorts the CC structure (for example, thickness reduction, volume reduction, envelop distortion and tissue alteration). It is important for the object detection module to be robust to these factors in order to successfully detect the CC. Hence, for improved CC detection accuracy, bag of freebies (BOF) augmentation technique is utilized in this study. Mosaic data augmentation method is used as a trainable BOF, that combines four training images into one in certain ratios without disturbing the object of interest (here, the CC). It also allows for the model to learn how to detect the object of interest at a smaller scale than the original scale. Additionally, we have also performed horizontal image flipping and gamma correction. Stochastic gradient descent is used as an optimizer with an initial learning rate of 0.01 and optimizer weight decay of 0.00005 per epoch. The YOLOv5 model is trained for 60 epochs with batch size of 2. Leaky-ReLU activation function is employed in middle layers and the sigmoid activation function is employed in the final detection layer. Binary cross entropy loss is used to compute the classification loss and the object-confidence loss. Additionally, complete intersection over union loss is used to compute the bounding-box localization loss \cite{b23}.

\subsection{Interpretibility using Eigen CAM}
Interpretability of any neural network models has reignited tremendously in recent years because of its ability to explain or interpret the "black-box" nature of deep neural architecture. In this regard, class activation maps (CAM) are widely used that can produce visual explanation to better understand the image classification problem. In our study, Eigen CAM technique \cite{b26} is utilized in order to explain the YOLOv5 predictions by visualizing principal components of learned representations from the last few convolutional layers of YOLOv5 architecture. Thus, Eigen CAM based coarse localization map is able to provide a visual representation of learned features of all layers of YOLO that stay relevant in the direction of maximum variation; which in turn help in identifying the most important regions in the CC, leading to the classification. The Eigen CAM is  class discriminative, computationally efficient, more consistent and robust against classification errors made by dense layers \cite{b26}. In Eigen CAM, the heatmap $\Theta_{Eigen CAM}$ is obtained by the projection of class activated output $\Phi_{L_{n}}$ on the first eigenvector ($V_{1}$) of $\Phi_{L_{n}}$ (Equation 1), where $\Phi_{L_{n}}$ is factorized using singular value decomposition. 
\begin{equation}
    \Theta_{Eigen CAM} = \Phi_{L_{n}}*V_{1} = W^{T}_{L_{n}}*I(i,j)
\end{equation}
The class activated output $\Phi_{L_{n}}$ is obtained by projecting the image I(i,j) onto the the last convolution layer $L_{n}$ and is computed by multiplying I(i,j) with the weight matrix of the last layer.

\begin{figure*}[htbp]
\centerline{\includegraphics[width=15cm]{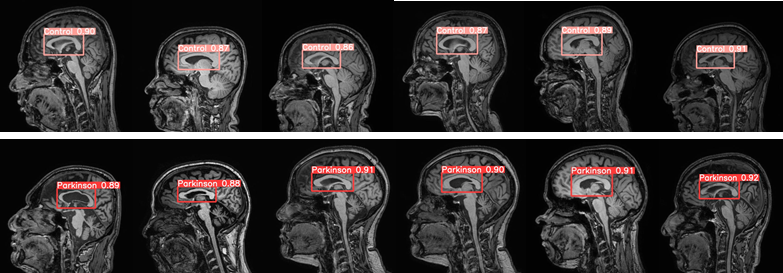}}
\caption{Automated detection of CC from MRI image volume using YOLOv5 with class score in case of HC (\textit{Top row}) and APD (\textit{Bottom row}).}
\label{detection}
\end{figure*}

\begin{figure*}[htbp]
\centerline{\includegraphics[width=15cm]{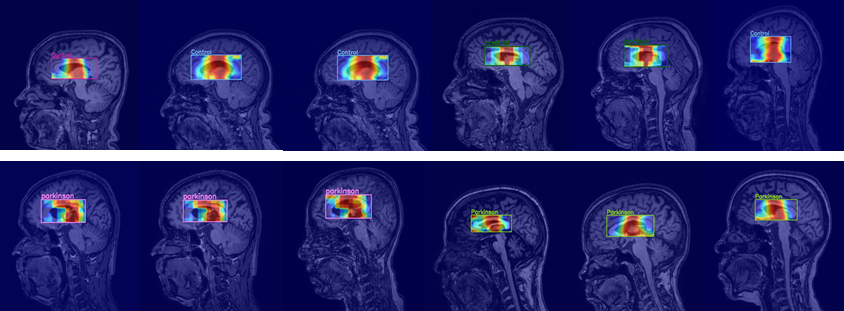}}
\caption{Eigen CAM interpretation on detected CC for HC (\textit{Top row}) and for APD (\textit{Bottom row}). Average class activation map shows the brightest location at mid-body of CC. This result is in-line with our previous publications \cite{b1,b2,b3} where we obtained the maximum callosal degeneration at its mid-body}
\label{cam}
\end{figure*}

\begin{figure}[htbp]
\centerline{\includegraphics[width=8.7cm]{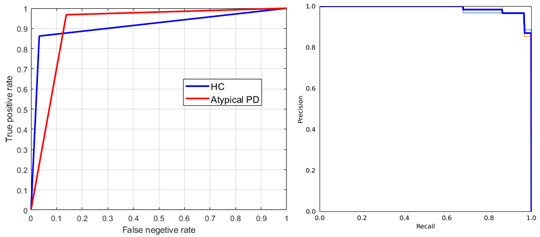}}
\caption{\textit{Left}: Receiver operator characteristics (ROC) curve of classification between HC vs. APD with respect to the result as tabulated in Table \ref{tab1}. Note the high degree of classification, with an area under the curve of 0.91 relating to both groups. Precision-Recall curve of the analysis is shown in the \textit{Right} column}
\label{fig1}
\end{figure}

\begin{table}[htbp]
\caption{Confusion matrix for one-round hold out validation along with classification performance using YOLOv5}
\begin{center}
\begin{tabular}{|c|c|c||c|c|c|c|c|}
\hline
\multicolumn{3}{|c||}{\textbf{Confusion matrix}} &\multicolumn{4}{|c|}{\textbf{Performance measures}}\\
\hline
& \textbf{HC} & \textbf{APD} & \textbf{Precision} & \textbf{Recall} & \textbf{F-score} & \textbf{Accuracy}\ \\
\hline
\textbf{HC} &27 &2 &0.95	&0.86	&0.91  &92\%  \\
\cline{1-6} 
\textbf{APD} &1 & 30 &0.88	&0.96	&0.92  &\\
\hline
\end{tabular}
\label{tab1}
\end{center}
\end{table}

\begin{table*}[htbp]
\caption{Comparison of the proposed methodology with our previously published works that used the same NIMHANS dataset for quantifying CC degeneration in MSA and PSP, showing an improvement of 5\% in classifying APD Vs. HC}
\begin{center}
\begin{tabular}{|c|c|c|c|c|c|}
\hline
\textbf{No.} &\textbf{Methods} & \textbf{Features} & \textbf{Analysis} &\textbf{Outcomes (Mean} &\textbf{Inference from study} \\
 & &  &  &\textbf{accuracy/p-value)} &\textbf{(A comparison across CC sub-regions)} \\
\hline
1 & CC Texture  &GLCM &SVM (rbf)  &82.4 \% & Highest textural alteration with highly significant   \\
\cline{1-1}
\cline{3-3}
\cline{5-5}
2 & Analysis \cite{b1,b2}  &Uniform LBP &classification  &87.4\%  & p-value (p$<<$0.001) was found consistently at \\
\cline{1-1}
\cline{3-3}
\cline{5-5}
3  &  &O-GDFB &  &69.2\% & mid-callosal regions, across all subjects\\
\hline
4 & CC Morphometry  & Thickness Profile & Statistical 
 & \multicolumn{2}{c|}{Maximum structural degeneration (in case of APDs) with highly significant} \\
\cline{1-1}
\cline{3-3}
5 &Analysis \cite{b3} & CC Area & testing & \multicolumn{2}{c|}{p-value (p$<<$0.0001) was obtained at CC mid-body regions, across all subjects} \\
\hline
6   &\textbf{Proposed work} &\multicolumn{2}{|c|}{\textbf{YOLOv5 + Eigen CAM}}
  &\textbf{92\%} & \textbf{Brightest location in heatmap occurs at CC mid-body}\\
\hline
\end{tabular}
\label{tab2}
\end{center}
\end{table*}

\section{Results and Discussion}

In this work, we have shown the ability of YOLOv5 framework to detect the CC structure in MRI for classifying APD Vs.HC. Subsequently, Eigen CAM is employed in order to explain the classification result. The result of this study is also compared against SOTA methodologies. All codes, results and intermediate results will be available in the following link:
https://github.com/kancharlavamshi/Yolov5forPD

\subsection{Detection of CC using YOLOv5}

YOLOv5 network is trained to detect CC from whole MRI. For each subject, a total of eight best MRI slices are chosen from sagittal plane where CC structure is well visible. This yields a total 320 MRI images. Due to smaller size of the dataset, 3 rounds of hold-out validation is performed. Datasets are randomly split into train, test sets, where 80\% of the data is used for training and 20\% data is used for testing. No data leakage is allowed between train and test sets. With YOLOv5, an average classification accuracy of 92\% was obtained. The result outperforms the SOTA methods that utilized the same dataset and CC morphometry and texture analysis for classification of PD vs. HC. The result of simultaneous localization of CC and classification of PD and HC is shown in Figure~\ref{detection} and tabulated in Table~\ref{tab1}.

\subsection{Eigen CAM identifies mid-callosal region as radiological marker}

Apart from reporting the classification accuracy as obtained from YOLOv5, we also illustrated the classification performance by Eigen CAM explanations. Representative images are randomly chosen from test split for illustrating CAM interpretation on YOLOv5. Figure~\ref{cam} shows that results of class activation map for CC input images of HC and patients suffering from APD. The average CAM activation is computed for identifying the key callosal sub-region that yield mean accuracy of 92\% in classifying HC vs. APD. The visualization from Eigen CAM highlights the mid-callosal region as a region which is most responsible for the obtained classification. It is observed that the heatmap is brightest at mid-body of CC which is found to be true for more than 90\% of images. This is in-line with our previously published works where the highest callosal tissue alteration was found at mid-body of CC in case of Parkinsonian disorders with respect to HC using CC morphometry and texture analysis \cite{b1,b2}. Axons of CC mid-body interconnect the areas of the motor cortex, responsible for preparation and sensory guidance of movement. The present knowledge of callosal involvement in neurodegenerative disorders suggest that in case of PDs, the changes in motor cortical activation are more pronounced which increases with disease severity (as in case of APDs). Moreover, as seen from Figure~\ref{cam}, along with mid callosal region, the brightest location in the heatmap also focus on the neighborhood of CC mid-body region (particularly, septum pellucidum, fornix and thalamus). This is quite obvious since, damage or abnormalities in any specific sub-cortical region (here, the CC) will have an effect in its surrounding areas. \par

To the best of our knowledge, this is the first study that uses interpretable Eigen CAM in combination with YOLOv5 deep learning framework in order to detect brain CC for classifying APD from HC. Comparison of this work with our previously published works on same dataset is presented in Table~\ref{tab2}. As seen from the table, the classification result of the present work outperforms our previous works, with an improvement of 5\% with respect to the average accuracy, obtained while classifying APD from HC.
There exists several other PD related research works that have been done in past few decades, aiming to classify PD Vs.HC and the results of many of these works are already summarized in literature by Mei.et.al \cite{b27}. Although some of these works achieved comparatively better accuracy in classifying PD vs. HC, the authors used different categories of data (eg. speech, movement data and handwritten patterns), image modalities and analyzed other sub-cortical regions with different techniques for classification. Contrary to all these works, in the current study, we have investigated a specific brain sub-cortical structure, the CC, due to its high relevance in neurodegeneration in PDs, using YOLOv5. Additionally, we have explained the YOLOv5 result using Eigen CAM to further identify the specific callosal sub-region, responsible for classification. The finding of this study demonstrates the possible utility of the proposed methodology in computer-aided diagnosis differential diagnosis of Parkinsonian disorders.

\section{Conclusion}

In this study, we have shown the potential of YOLOv5 for automated detection of brain CC to classify APD and HC. The result of classification is further explained using Eigen CAM and compared against our previous results, which utilized callosal morphometry and texture analysis, on the same dataset. In the proposed method, we have obtained an improvement of 5\% in classifying APD vs. HC. The availability of limited dataset is one limitation of this study. Moreover, it is necessary to fine tune the model parameters and to use other CAM versions for further improvisation. Nevertheless the results of our analysis showed its potential for future studies of computer aided differential diagnosis of Parkinsonian disorders using CC.




\begin{thebibliography}{00}
\bibitem{b1} Bhattacharya D, Sinha N, Saini J et.al. (2020). A New Statistical Framework for Corpus Callosum Sub-Region Characterization Based on LBP Texture in Patients With Parkinsonian Disorders: A Pilot Study. Front. Neurosci. 14:477. doi: 10.3389/fnins.2020.00477

\bibitem{b2} D. Bhattacharya, S. K. Vengalil, N. Sinha, J. Saini et.al (2019). "Structural MRI based texture analysis of corpus callosum in patients with Progressive Supraneuclear Palsy. IEEE Region 10 Conference (TENCON), pp. 441-446, doi: 10.1109/TENCON.2019.8929403.

\bibitem{b3} D. Bhattacharya, N. Sinha, J. Saini (2017). Quantitative analysis of structural variations in corpus callosum in adults with multiple system atrophy (MSA). Medical Imaging 2017: Biomedical Applications in Molecular, Structural, and Functional Imaging, vol. 10137. https://doi.org/10.1117/12.2253873

\bibitem{b4} Saini J., Mangalore S, Lenka A, Pal PK et.al (2017). Role of corpus callosum volumetry in differentiating the subtypes of progressive supranuclear palsy and early parkinson’s disease. Movment Disorder Clinical Practice, 4(4):552–558. doi: https://doi.org/10.1002/mdc3.12473.

\bibitem{b5} Reneta Mileva-Daniel J. A., Connolly Patricia E., Stavros M. Stivaros et.al (2016). Quantification of structural changes in the corpus callosum in children with profound hypoxic–ischaemic brain injury. Pediatric Neuroradiology, 46(1):73–81. doi: https://doi.org/10.1007/s00247-015-3444-3.

\bibitem{b6} El-Baz A. Elnakib A.-Giedd J. Rumsey J. et al. Casanova, M. F.(2010). Corpus callosum shape analysis with application to dyslexia. Translational Neuroscience, 1(2):124–130. doi: https://doi.org/10.2478/v10134-010-0017-8.


\bibitem{b8} Narr-K. L. Philips O. R. Nuechterlein K. H. Asarnow R. F. Toga A.W. et al. Joshi, S. H. (2013) Statistical shape analysis of the corpus callosum in schizophrenia. Neuroimage, 64:547–559. doi: https://doi.org/10.1016/j.neuroimage.2012.09.024

\bibitem{b9} Wang X.-Gao W. Shi C. Ge H. Shen H. et al. Zhu, M. (2014). Corpus callosum atrophy and cognitive decline in early alzheimer’s disease: longitudinal MRI study. Dementia and Geriatric Cognitive Disorder, 37(3-4):214–222. doi:
https://doi.org/10.1159/000350410.


\bibitem{b11} Bachman A. H. Lee S. H. Sidtis J. J. Ardekani B. A. Elahi, S. and Alzheimer’s Disease Neuroimaging Initiative (2015). Corpus callosum atrophy rate in mild cognitive impairment and prodromal alzheimer’s disease. Journal of Alzheimers Disease, 45(3):921–931. doi: https://doi.org/10.3233/JAD-142631.



\bibitem{b14} Elke Hattingen Patrick, Jung Oliver Singer et.al (2007). Human motor corpus callosum: Topography, somatotopy, and link between microstructure and function. The Journal of Neuroscience, page 12132–12138. doi: https://doi.org/10.1523/JNEUROSCI.2320-07.2007.

\bibitem{b15} Karl Sjostrand Egill, Rostrup Frederik Barkhof et.al (2006). Corpus callosum partitioning schemes and their effect on callosal morphometry. In Proc. International Society of Magnetic Resonance In Medicine - ISMRM 2006, Seattle, Washington, USA.


\bibitem{b17} Brusini I, Platten M, Ouellette R, Piehl F, Wang C, Granberg T (2022). Automatic deep learning multicontrast corpus callosum segmentation in multiple sclerosis. J Neuroimaging. 2022 May;32(3):459-470. doi: 10.1111/jon.12972.


\bibitem{b19} Park G, Kwak K, Seo SW and Lee J-M (2018). Automatic Segmentation of Corpus Callosum in Midsagittal Based on Bayesian Inference Consisting of Sparse Representation Error and Multi-Atlas Voting. Front. Neurosci. 12:629. doi: 10.3389/fnins.2018.00629

\bibitem{b20} Anjali Chandra, Shrish Verma, A.S. Raghuvanshi, Narendra Kuber Bodhey (2022). CCsNeT: Automated Corpus Callosum segmentation using fully convolutional network based on U-Net, Biocybernetics and Biomedical Engineering, Volume 42(1), Pages 187-203. doi: https://doi.org/10.1016/j.bbe.2021.12.008.



\bibitem{b23} Ziyu Zhao, Xiaoxia Yang et.al (2021). Real‑time detection of particleboard surface defects based on improved YOLOV5 target detection. Scientific Reports. 11:21777. doi: /10.1038/s41598-021-01084-x

\bibitem{b24} Qingqing Xu, Zhiyu Zhu et. al (2021). Effective Face Detector Based on YOLOv5 and Superresolution Reconstruction. Computational and Mathematical Methods in Medicine. Vol. 2021, Article ID 7748350. doi: https://doi.org/10.1155/2021/7748350

\bibitem{b25} Renjie Xu, Haifeng Lin et. al (2021). A Forest Fire Detection System Based on Ensemble Learning. Forests. 12, 217. doi: https://doi.org/10.3390/f12020217

\bibitem{b26} M. B Muhammud and M. Yeasin (2021). Eigen‑CAM: Visual Explanations for Deep Convolutional Neural Networks. SN Computer Science. doi: https://doi.org/10.1007/s42979-021-00449-3

\bibitem{b27} Jie Mei, Christian Desrosiers and Johannes Frasnelli (2021). Machine Learning for the Diagnosis of Parkinson’s Disease: A Review of Literature. Front. Aging Neurosci. 13:633752. doi: 10.3389/fnagi.2021.633752




\end{thebibliography}
\end{document}